\documentclass[floats,floatfix,showpacs,amssymb,prd,twocolumn,superscriptaddress,nofootinbib,reprint]{revtex4-2}

\usepackage{amssymb,amsmath,verbatim,mathtools,needspace,enumitem,etoolbox,graphicx,physics,microtype,afterpage,xspace,tabularx,lmodern,multirow}
\usepackage{siunitx}
\usepackage{gensymb}
\usepackage[dvipsnames, usenames]{xcolor}
\definecolor{linkcolor}{rgb}{0.0,0.3,0.5}
\usepackage[unicode, colorlinks=true, linkcolor=linkcolor, citecolor=linkcolor, filecolor=linkcolor, urlcolor=linkcolor, linktocpage, breaklinks]{hyperref}
\usepackage[all]{hypcap}
\usepackage[T1]{fontenc}
\usepackage[utf8]{inputenc}
\usepackage[usenames,dvipsnames]{xcolor}
\hypersetup{colorlinks=true,citecolor=romared,linkcolor=romared,urlcolor=romared}

\definecolor{romared}{RGB}{142,0,28}

\newcommand{\be}{\begin{equation}}
\newcommand{\ee}{\end{equation}}

\def\be{\begin{equation}}
\def\ee{\end{equation}}
\newcommand{\beq}{\begin{eqnarray}}
\newcommand{\eeq}{\end{eqnarray}}

\usepackage{aas_macros}
\usepackage{makecell}
\usepackage{soul}

\usepackage{lipsum}
\usepackage{orcidlink}

\newcommand{\hac}[1]{{\textcolor{black}{{{#1}}}}}

\newcolumntype{Y}{>{\centering\arraybackslash}X}

\begin{document}
\title{21-cm fluctuations from primordial magnetic fields}

\author{Hector Afonso G. Cruz \,\orcidlink{0000-0002-1775-3602}}
\email{hcruz2@jhu.edu}
\affiliation{William H. Miller III Department of Physics and Astronomy, Johns Hopkins University, 3400 N. Charles Street, Baltimore, Maryland, 21218, USA}

\author{Tal Adi\,\orcidlink{0000-0002-5763-9353}}
\email{talabadi@post.bgu.ac.il}
\affiliation{Department of Physics, Ben-Gurion University of the Negev, Be’er Sheva 84105, Israel}

\author{Jordan Flitter\,\orcidlink{0000-0001-8092-8228}} 
\email{jordanf@post.bgu.ac.il}
\affiliation{Department of Physics, Ben-Gurion University of the Negev, Be’er Sheva 84105, Israel}

\author{Marc Kamionkowski \,\orcidlink{0000-0001-7018-2055}}
\email{kamion@jhu.edu}
 \affiliation{William H. Miller III Department of Physics and Astronomy, Johns Hopkins University, 3400 North Charles Street, Baltimore, Maryland, 21218, USA}

\author{Ely D. Kovetz\,\orcidlink{0000-0001-9256-1144}} 
\email{kovetz@bgu.ac.il}
\affiliation{Department of Physics, Ben-Gurion University of the Negev, Be’er Sheva 84105, Israel}

\begin{abstract}
The fluid forces associated with primordial magnetic fields (PMFs) generate small-scale fluctuations in the primordial density field, which add to the $\mathrm{\Lambda CDM}$ linear matter power spectrum on small scales. These enhanced small-scale fluctuations lead to earlier formation of galactic halos and stars and thus affect cosmic reionization. We study the consequences of these effects on 21-cm observables using the semi-numerical code {\tt\string 21cmFAST v3.1.3}. We find the excess small-scale structure generates strong stellar radiation backgrounds in the early Universe, resulting in altered 21-cm global signals and power spectra commensurate with earlier reionization. We restrict the allowed PMF models using the CMB optical depth to reionization. Lastly, we probe parameter degeneracies and forecast experimental sensitivities with an information matrix analysis subject to the CMB optical depth bound. Our forecasts show that interferometers like HERA are sensitive to PMFs of order $\sim \mathrm{pG}$, \hac{nearly an order of magnitude stronger than existing 21cm observatories and future standalone CMB experiments}.

\end{abstract}

\date{\today}
\maketitle



\section{Introduction}\label{sec:intro}

The existence of magnetic fields at a diversity of astrophysical scales present a captivating, yet mysterious, puzzle piece of cosmic evolution. Observations of quasar spectra \citep{bernet08} and far-infrared radio correlation studies \citep{murphy09} have shown galaxies to exhibit $10^{-5} \, \si{G}$ fields up to redshifts of $z \lesssim 3$  \citep{beck96, beck13}. Other probes of galaxy clusters show fields coherent on $10 - 50 \, \si{kpc}$ scales \citep{beck12} at strengths of $\sim 10^{-6} \, \si{G}$ \citep{clarke01, govoni04, vogt05, dubois08, ryu08}, which fall to $\sim 10^{-8} \, \si{G}$ \citep{miniati11} in filaments. Magnetic fields at even larger scales and in cosmological contexts remain enigmatic. Gamma ray studies imply the existence of sub-$\si{nG}$ level fields \citep{neronov10, tavecchio10, taylor11, broderick12} with coherence lengths of $0.1-1 \, \si{Mpc}$ \citep{kronberg94}. 

The origins of such large-scale fields are topics of persisting study (see Refs.~\citep{widrow02, yamazaki12, durrer13, subramanian16} for recent reviews.) Field sourcing from plasma instabilities and galactic winds \citep{miniati11b, schlickeiser03, bertone06}, from Biermann battery-type mechanisms \citep{biermann50}, from primordial black hole disks \citep{papanikolaou23a, papanikolaou23b}, from the first stars \citep{ando10, schleicher10, sur10}, or from magnetogenesis during inflation or early phase transitions \citep{turner88, ratra92, grasso01, giovannini04, ashoorioon05, kandus11} have been posited. Dynamo processes are thought to play a role in cosmic magnetic field amplification and evolution \citep{federrath11, schober12}, perhaps regulated by turbulence in young galaxies \citep{schober13a} or the collapse of primordial haloes \citep{schober13b}. Although the models generating these large-scale weak fields are favored to be primordial in origin, none of these prototypes have empirical confirmation.

Primordial magnetic fields (PMFs) have a variety of consequences on cosmological and astrophysical observables. Magnetized gas in the universe may affect passing CMB photons and induce a frequency-dependent Faraday rotation of CMB polarization \citep{scoccola04, kosowsky05, giovannini08, kahniashvili09, pogosian11, kahniashvili10, yamazaki10, kunze15}. Heating of the intergalactic medium (IGM) from ambipolar diffusion and the decay of magnetohydrodynamic (MHD) turbulence can occur, which may raise filtering masses \citep{gnedin00a, gnedin00b}, hinder halo collapse \citep{sethi05, sethi08, schleicher08}, impact the UV luminosity function of early galaxies \citep{schleicher11}, and alter the cosmic neutral gas fraction \citep{chluba15}. 

The most prominent consequence of PMFs is the generation of small-scale matter density perturbations \citep{wasserman78, coles92} at early times. PMF inhomogeneities involve energy-density and pressure fluctuations that move the primordial plasma to which magnetic fields are pinned. At recombination, these perturbations induced in the baryon density---and thus the total matter density---result in excess power in the primordial mass distribution on scales of $10^1 \lesssim k  \lesssim 10^{3} \ \left( \si{h^{-1} \, Mpc}\right)$ \citep{kim96, gopal03, shaw10, shaw12, kunze21, kunze22}. Such early small-scale baryon inhomogeneities from magnetic fields of strength $\sim 0.1 \, \mathrm{nG}$ can increase the average recombination rate and reduce the sound horizon at recombination, thus providing a promising solution to the Hubble tension \citep{jedamzik20, neronov21, divalentino21, jedamzik21, galli22, hosking22, lucca23a, lucca23b, li23, jedamzik23}. 

Of paramount interest in this study are magnetic effects after decoupling, where PMF-induced perturbations affect the density of small haloes at high redshift and the Thomson scattering optical depth \citep{sethi09, kahniashvili13}. Increased mini-halo abundances can affect the Lyman-$\alpha$ forest in quasar absorption spectra leading to increased effective Lyman-$\alpha$ opacities \citep{pandey13, chongchitnan14, pandey15}. Amplified small-scale structure can also boost star formation rates from dwarf galaxies at early times \citep{sanati20}, which could alleviate current constraints on the ionizing photon budget at high redshifts and solve the photon-starvation problem \citep{miralda03, meiksin05, bolton07}. Cumulatively, these effects are postulated to significantly impact the post-recombination growth of structure, the ionization history of neutral hydrogen, and the thermal evolution of the IGM \citep{tashiro06a}. 

A study of PMF signatures during Cosmic Dawn (CD) and the Epoch of Reionization (EoR), most notably with 21-cm line-intensity mapping (LIM), should reveal how these puzzle pieces fit. Disentangling the 21-cm signal involves solving a multi-scale problem, in which inter-atomic processes of neutral hydrogen spin-flip transitions can cause large-scale observables. Mapping 21-cm fluctuations can assess the aggregate impact of the first emitting sources on the IGM \citep{furlanetto06}, constrain reionization properties \citep{loeb04} and the growth of cosmic structure \citep{furlanetto06b}, and probe fundamental physics \citep{pritchard12, barkana05}. Specifically, enhanced small-scale structure can imprint on 21-cm fluctuations \citep{furlanetto06c}, which has been studied in the case of magnetically-induced perturbations \citep{tashiro06b, schleicher09, shiraishi14, kunze19a, kunze19b, minoda22, kunze23}.

In this paper we study the prospects of probing PMFs with forthcoming 21-cm measurements. We use existing prescriptions to model the magnetically-induced matter power spectrum (\citep{magFieldPaper}, see Refs.~\citep{kim96, wasserman78} for earlier work.) We employ a modified version of {\tt\string 21cmFAST v3.1.3} \citep{murray20, mesinger11} to model the 21-cm global signal and power spectrum. We then evaluate the plausibility of different PMF cosmologies in three steps. Firstly, we restrict the parameter space of allowed magnetic models by using external constraints on their ionization histories. Subsequently, we study the credibility of these restrictions by studying the degeneracies in our astrophysical and cosmological assumptions. Through an information matrix analysis, we probe the covariance in our chosen simulation parameters subject to bounds from the optical depth to reionization inferred from the CMB \citep{debelsunce21}. Finally, we see how these bounds compare to what 21-cm experiments can detect. We marginalize over all non-PMF parameters to forecast sensitivities for of the upcoming \textit{Hydrogen Epoch of Reionization Array} (HERA) experiment \citep{deboer17}. \hac{We show that HERA is competitively stronger in constraining PMF models than existing 21cm observatories and next-generation standalone CMB experiments by nearly an order of magnitude.}

This work is organized as follows. In Sec. \ref{sec:matterPK} we denote our magnetic perturbation formalism in four steps. We introduce our PMF parameterization in \ref{sec:magFieldPar}, specify pre-recombination damping effects in \ref{sec:damp}, formulate the magnetically-induced matter power spectrum in \ref{sec:totalPK}, and explain post-recombination matter perturbation processing in \ref{sec:jeans}. In Sec. \ref{sec:21cm}, we highlight the relevant details in our 21-cm formalism. We show the results of our simulations in Sec. \ref{sec:sim}. In Sec. \ref{sec:ionHistory} we assess the PMF parameter space in agreement with the optical depth to reionization obtained from CMB polarization \citep{debelsunce21}. We show our information matrix forecast in Sec. \ref{sec:fisher}, and forecast HERA sensitivities in \ref{sec:hera}. We conclude our work in Sec. \ref{sec:conclusion}.

\section{Magnetically-Induced Perturbation Formalism} \label{sec:matterPK}

Here we present our parameterization of the magnetic field, describe its evolution with time, and detail the density perturbations that it induces. We work in SI units, with $\mu_0$ being the magnetic permeability of the vacuum, and the energy density in the magnetic field is $\rho_B=\textbf{B}^2/(2\mu_0)$.

\subsection{Primordial Magnetic Field Parameterization} \label{sec:magFieldPar}

We begin with a magnetic field ${\bf B}({\bf x},t)$ of primordial origin (i.e. either generated during inflation or some time well before recombination), as a function of comoving position ${\bf x}$ and time $t$. From \citep{wasserman78, kim96, fedeli12}, these fields evolve as
\begin{equation}
    a^2(t){\bf B}({\bf x},t) = {\bf B}({\bf x},t_0) \equiv {\bf B_0}({\bf x}),
    \label{eq:bFieldEvol}
\end{equation}
where $a(t)$ is the scale factor, $t_0$ is the age of the universe, and quantities with a subscript 0 are evaluated at the present time. The magnetic field ${\bf B}({\bf x},t)$ is itself a realization of a random field with power spectrum $P_B(k,t)$ given by
\begin{equation}
    \left\langle\tilde{B}_i(\boldsymbol{k}, t) \tilde{B}^{*}_j(\boldsymbol{k}^{\prime}, t)\right\rangle=(2 \pi)^3 \delta_{D}\left(\boldsymbol{k}-\boldsymbol{k}^{\prime}\right)\frac{P_{ij}}{2} P_B(k, t),
\end{equation}
where $\tilde{\bf B}({\bf k},t)= \int   {\bf B}({\bf x},t) \, e^{-i {\bf k} \cdot {\bf x}} \, d^3x$ is its Fourier transform, and $P_{ij} = \left(\delta_{i j}-k_i k^{\prime}_j/k^2\right)$ is the projection tensor to enforce a divergence-free magnetic field. The primordial magnetic-field power spectrum itself can be described by a simple power-law of the form
\begin{equation}
P_{B, \mathrm{prim}}(k,t) = A_B(t) k^{n_B},
\end{equation}
defined by some amplitude $A_B(t)$ and magnetic spectral index $n_B$. It is common to consider only $n_B > -3$, as infrared divergences appear at lower values of the magnetic spectral index. By convention, the normalization of the magnetic power spectrum is referenced in terms of the magnetic-field variance, smoothed using a real-space Gaussian filtering function on a comoving scale of $\lambda = 1 \, \si{Mpc}$ defined as
\begin{align} \label{eq:sigmaB}
\sigma_B^2(\lambda, t) &=\frac{1}{2 \pi^2} \int_0^{\infty} k^2 P_{B, \mathrm{prim}}(k, t) \exp \left( -k^2 \lambda^2 \right)\mathrm{~d} k \nonumber\\
& =\frac{A_B(t)}{\left( 2\pi \right)^2 \lambda^{3+n_B}} \Gamma \left(\frac{ n_B+3}{2} \right),
\end{align}
which exhibits a time dependence that follows $\sigma_B(\lambda, t) = \sigma_B(\lambda, t_0)/a^2(t)$ from that of Eq.~(\ref{eq:bFieldEvol}). We therefore characterize PMFs via a two-parameter model using the magnetic spectral index $n_B$ and the comoving smoothed amplitude $\sigma_{B,0} \equiv \sigma_B \left(\lambda = 1 \, \mathrm{Mpc}, t = t_0\right)$.

\hac{It is important to note that both the magnetic-field power spectrum and the matter overdensities it sources are subject to scale-dependent damping processes. As such, both are defined within the scale bounds $k_\mathrm{min} \leq k \leq k_\mathrm{max}$. As magnetic fields are assumed to be correlated at very large scales, we set $k_\mathrm{min} = 0$. The upper bound $k_\mathrm{max}$ requires a more complex treatment. It is convention to apply a strict cutoff in both magnetic fields and the induced matter power spectrum at a scale $k_\mathrm{max}$ of choice. To avoid a sharp "sawtooth"-shaped power spectrum, we instead employ a more physically-motivated suppressive formalism involving two damping processes, to be detailed in subsequent sections.}

\subsection{Pre-Recombination Alfv\'en Damping} \label{sec:damp}

\begin{figure}[t!]
    \centering
    \includegraphics[width=\textwidth/2]{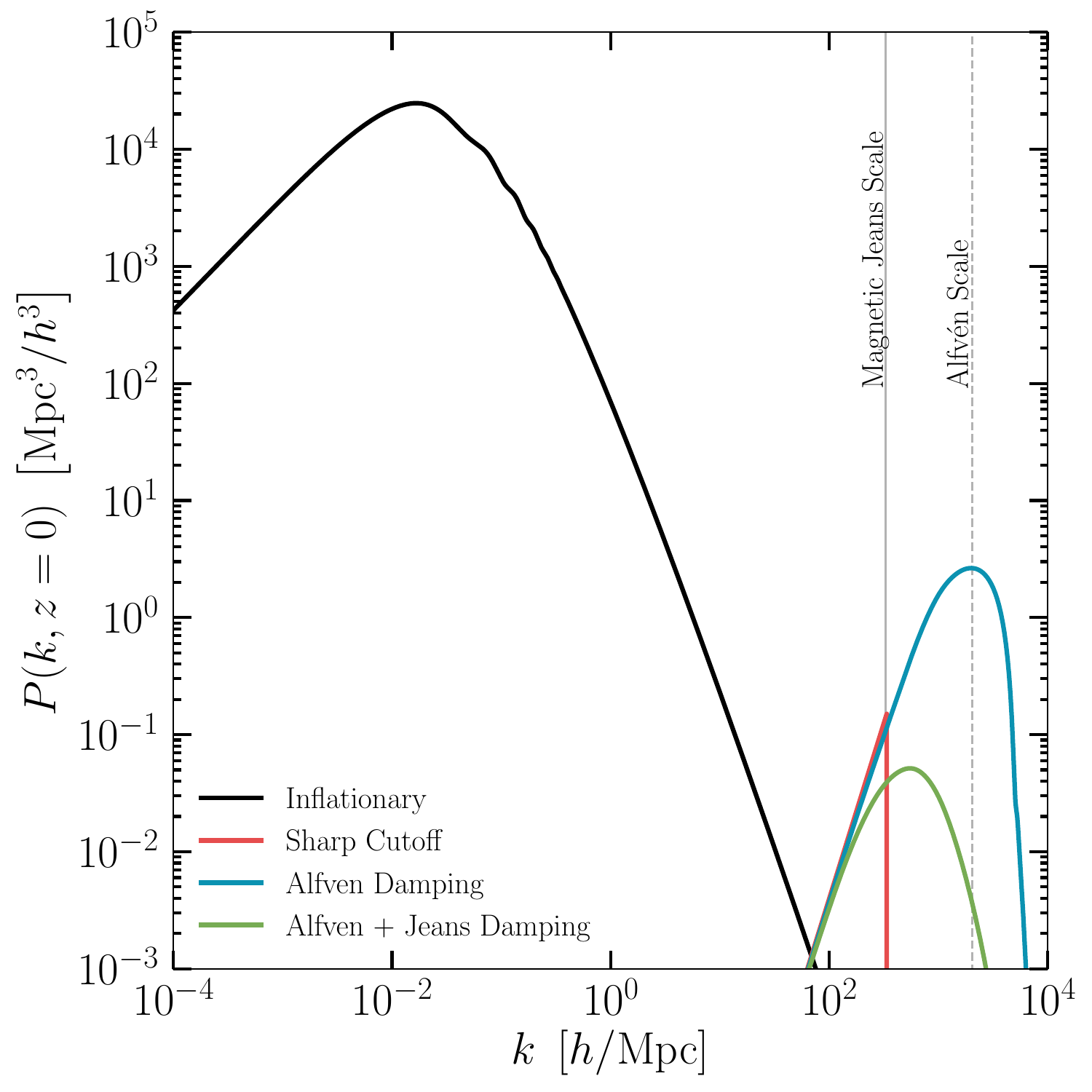}
    \caption{To demonstrate the different damping procedures outlined in Sec. \ref{sec:damp}, we plot an example PMF-induced matter power spectrum $\Pi(k)$ overplotted onto the $\mathrm{\Lambda CDM}$ linear matter power spectrum (black). In red is the computed $\Pi(k,z = 0)$ with parameters $n_B = -2.0, \sigma_{B,0} = 0.01 \, \mathrm{nG}$, exhibiting no damping and a non-physical hard cutoff at $k_\mathrm{max} = k_B$. In blue, we plot a power spectrum incorporating the exponential decay terms of Alfv\'en damping due to radiative viscosity. The green plot incorporates both Alfv\'en and Jeans damping. For reference, in translucent solid gray is the magnetic Jeans scale $k_B$ and in translucent dashed gray is the Alfv\'en scale $k_A$ for this specific PMF model. \hac{Though it is conventional to use the sharp cutoff approach (red) at $k = k_B$, we employ the more physically-motivated approach of including both suppression effects (green) in every computed $\Pi(k)$ used for our 21cm analysis.} }
  \label{fig:damp}
\end{figure}

A crucial hydrodynamical effect that occurs prior to last scattering is the Alfv\'en damping of early magnetic fields. Alfv\'en damping is a phenomenon in which magnetic fields are dissipated due to radiative viscosity \citep{jedamzik98, subramanian98}. Early magnetic fields interact with primordial plasma to produce Alfv\'en waves whose characteristic velocity is
\begin{equation}
    v_A^2(t) \equiv \frac{\sigma_B^2 ( \lambda_A ,t )}{\mu_0 \left(\rho_\mathrm{tot}(t) + p_\mathrm{tot}(t) \right)}, \label{eq:alfvel}
\end{equation}
where $\rho_\mathrm{tot}(t) + p_\mathrm{tot}(t) \equiv \rho_r + p_r + \rho_b + p_b = \rho_b + 4 \rho_r / 3 $ is the total density and pressure of baryons and relativistic species. The Alfv\'en scale $\lambda_A = 2 \pi / k_A$ is the characteristic scale below which viscous damping becomes predominant, described by
\begin{align}
    \frac{1}{k_A^2} & \equiv \int_0^{t_\mathrm{rec}} \frac{v_A^2(t) \tau_c(t)}{a^2(t)} \, dt, \nonumber \\ \label{eq:alfvenScale}
\end{align}
where $\tau_c^{-1}(t) \equiv c n_e(t) \sigma_T$ is the timescale for Thomson scattering and $t_\mathrm{rec}$ is the time of recombination. We use {\tt\string HyRec-2} \citep{lee20, ali11} to compute the free-electron density $n_e(t) = x_e(t) n_\mathrm{H}(t)$. It is essential to note that the magnetic amplitude $\sigma_B(\lambda_A,t)$ in Eq.~(\ref{eq:alfvel}) differs from the $1 \, \mathrm{Mpc}$ smoothed magnetic amplitude used in this work due to the definition in Eq.~(\ref{eq:sigmaB}); these are related by
\begin{equation} 
    \sigma_B^2\left(\lambda_A \right) = \sigma_B^2\left(\lambda = 1 \, \mathrm{Mpc}\right) \left( \frac{1 \, \mathrm{Mpc}}{\lambda_A}\right)^{3 + n_B}.
\end{equation}
Evaluating the integral in Eq.~(\ref{eq:alfvenScale}), one gets the approximation
\begin{equation}
    k_A \approx \left(   \frac{\sigma_{B,0}^2 /\mathrm{nG}^2 }{(2 \pi)^{3+n_B} \cdot 4.2 \times 10^5}  \right)^{-1/(5+n_B)} \, \mathrm{Mpc^{-1}}.
\end{equation}
It can be shown \citep{shaw10, shaw12, saga18, minoda22} that magnetic fields are damped approximately as
\begin{equation}
    \tilde{B}_i(\boldsymbol{k}, t) = \tilde{B}_{0i}(\boldsymbol{k}, t) \exp \left( -k^2 / k_A^2 \right),
\end{equation}
and therefore, the resulting processed magnetic field power spectrum then becomes
\begin{equation}
    P_{B}(k,t) = A_B(t) k^{n_B} \exp \left( -2k^2 / k_A^2 \right),
\end{equation}
whose form we use in following sections.

\begin{figure*}[t!]
    \centering
    \includegraphics[width=\textwidth]{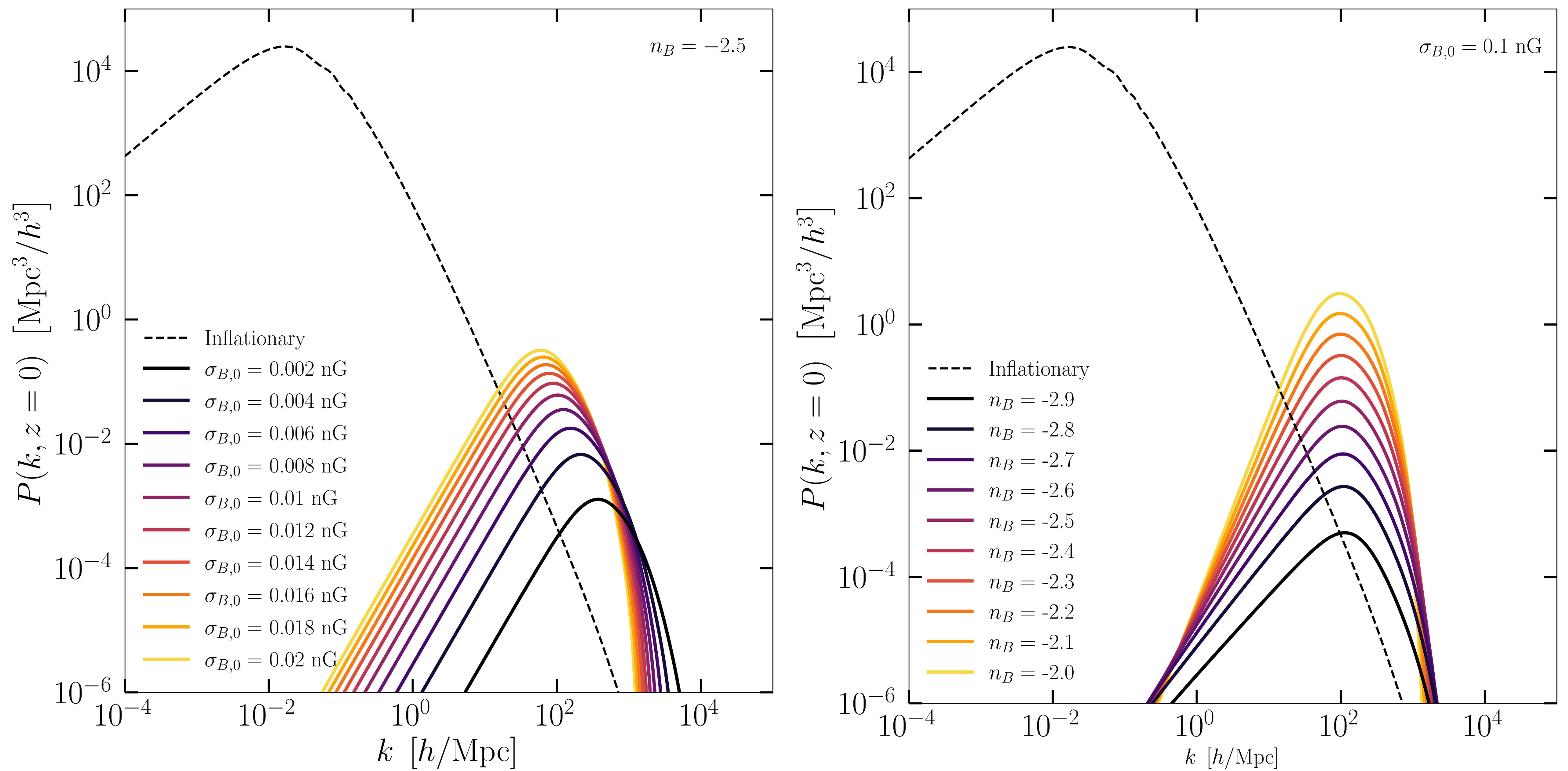}
    \caption{Present-day magnetic field-induced matter power spectra $\Pi_0(k) = M^2(t_0)\Pi(k)$ of different PMF models, with inflationary $\mathrm{\Lambda CDM}$ matter power spectrum overplotted in dashed black for reference. Left: PMF-induced $\Pi_0(k)$ at fixed $n_B = -2.5$ and varying amplitude $\sigma_{B,0}$. As $\sigma_{B,0}$ increases, $\Pi_0(k)$ increases, and its peak shifts towards smaller $k$; this is due to the magnetic Jeans length $\lambda_B$ increasing with $\sigma_{B,0}$ in Eq.~(\ref{eq:jeansLength}). Right: PMF-induced $\Pi_0(k)$ at fixed $\sigma_{B,0} = 0.1 \, \mathrm{nG}$ and varying magnetic spectral indices. Increasing $n_B$ steepens the slopes of $\Pi_0(k)$. In both plots, $\Pi_0(k)$ becomes dominant at scales $10^1 \lesssim k \lesssim 10^4 \ \mathrm{Mpc}/h$. We incorporate the total matter power spectrum $P_m(k)$ which involves a sum of the $\mathrm{\Lambda CDM}$ + a PMF matter power spectra into our inspection of 21-cm observables.}
  \label{fig:PK}
\end{figure*}

\subsection{Induced Matter Power Spectrum} \label{sec:totalPK}

As the Universe decouples at a redshift of $z \approx 1100$, radiation forces become subdominant to magnetic forces. Such remnant Lorentz pressures modify the evolution of structure and induce small-scale matter perturbations. The total matter power spectrum of inflationary and PMF-induced perturbations is given by
\begin{equation} \label{eq:totalMatterPK}
    P_m(k,t) = D_{+}^2(t)P_\mathrm{lin}(k) + M^2(t) \Pi(k),
\end{equation}
where $P_\mathrm{lin}(k)$ is the present-day linear matter power spectrum and $D_{+}(t)$ is the cosmological growth factor, normalized such that $D_{+}(t = t_0) \equiv 1$ at the present day. Finally, $M^2(t)\Pi(k) \, \si{[Mpc^4]}$ is the PMF-induced matter power spectrum, with the magnetic growth factor $M(t) \, \si{[s^4]}$ being the temporal evolution of PMF-induced matter perturbations assuming the null initial conditions $M(t_\mathrm{rec}) =\dot{M}(t_\mathrm{rec}) \equiv 0$ at recombination. According to the formalism developed in \citet{magFieldPaper}, the $\Pi(k)$ power spectrum can be evaluated by
\begin{align} 
    \Pi(k) &= \left(\frac { f_b k}{4\pi \mu_0 \rho_{b,0}}
     \right)^2 \int_0^\infty k_1^2\, dk_1
     \int_{-1}^1 d\mu\, \nonumber \\
    &\times  P_B(k_1) P_B\left(\sqrt{k^2+k_1^2 -2k k_1
     \mu} \right)\nonumber \\
    & \times  \left[k^2+(k^2-2 k k_1\mu)\mu^2 \right],
\label{eqn:PiPower}    
\end{align}
where $f_b \equiv \Omega_\mathrm{b,0} / \Omega_\mathrm{m,0}$ is the fraction of baryons to total matter, $\rho_{\mathrm{b,0}}$ is the present-day baryon density, and $\mu$ is the cosine of the angle between vectors $\mathbf{k}$ and $\mathbf{k_1}$. Lastly, the PMF-induced matter growth factor is the solution of the differential equation 
\begin{equation}
\ddot{M}(t)+2 H(t) \dot{M}(t)-4 \pi G \rho_{\mathrm{m}}(t) M(t)=\frac{1}{a^3(t)}.
\end{equation}
To streamline computations of the 21-cm signal, we make one critical simplification. Both the cosmological and magnetic growth factors normalized to the present time are nearly identical at low redshifts and differ by at most $\approx 7 \%$ by $z=30$. Due to the negligible difference, we make the approximation $M(t)/M(t_0)  \approx D_+(t)/D_+(t_0)$ when conducting our simulations.

\subsection{Post-Recombination Jeans Suppression} \label{sec:jeans}

As aforementioned in Sec.~\ref{sec:magFieldPar}, the magnetic-field power spectrum is defined within the bounds $k_\mathrm{min} \leq k \leq k_\mathrm{max}$. While the magnetic field upper bound $k_\mathrm{max}$ is the Alfv\'en scale, the upper bound on PMF-generated matter perturbations is customarily taken to be the magnetic Jeans length $\lambda_B = 2 \pi/k_B(t)$, below which magnetic field pressure gradients hinder the gravitational collapse of perturbations. The comoving magnetic Jeans scale grows with increasing magnetic field strength, described by
\begin{align}
    \label{eq:jeansLength}
    \lambda_B &= \left[ \frac{16\pi}{25}\frac{\sigma_{B,0}^2}{\mu_0 G \rho_{m,0} \rho_{b,0}}\lambda^{3+n_B} \right]^{1/\left(5+n_B\right)} \nonumber \\
    &\approx \left( 0.22  \left(\frac{\lambda}{\mathrm{Mpc}}\right)^{3+n_B} \frac{\sigma_{B,0}^2}{\mathrm{nG^2}}\right) ^{1/\left(5+n_B\right)}  \ \mathrm{Mpc},
\end{align}
in terms of Newton's constant $G$, the typical convolution length $\lambda = 1 \, \mathrm{Mpc}$ of the magnetic amplitude, and fundamental constants.

Magnetic Jeans instabilities suppress matter perturbations on scales near the Jeans length \citep{bi92, bi97}. The PMF-induced matter perturbations that form first are baryonic and are therefore subject to magnetic Jeans criteria; these initial baryonic gravitational potentials will eventually attract dark matter. To first order, the initial baryonic perturbations follow a Jeans damping factor $\delta_\mathrm{b} = \delta_\mathrm{cdm}/ \left( 1 + k^2 / k_B^2 \right)$ \citep{gnedin98, gnedin00a, zaroubi06, pandey13, kulkarni15}, which has shown considerable agreement with baryonic simulations of high-redshift galaxies and minihaloes \citep{naoz05, naoz07, naoz09, naoz11, naoz13}. Consequently, we assume that the resulting PMF-induced matter power spectrum $\Pi(k)$ is subject to a Jeans modification
\begin{equation}
\Pi(k) \rightarrow \frac{\Pi(k)}{\left( 1 + k^2/k_B^2\right)^2}.
\end{equation}

We portray the aggregate contributions of Alfv\'en and Jeans suppressive effects on the PMF-induced matter power spectrum in Fig.~\ref{fig:damp}, for given parameters $n_B = -2.0, \sigma_{B,0} = 0.01 \, \si{nG}$. Across all PMF model parameters, $k_B < k_A$; Alfv\'en damping affects magnetic fields at scales smaller than the magnetic Jeans limitations on the collapse of perturbations. \hac{While Fig.~\ref{fig:damp} shows that Jeans suppression has a much stronger effect than Alfv\'en damping on the eventual PMF-induced matter power spectrum, we caution that Alfv\'en damping is not negligible. Alfv\'en and Jeans suppression effects damp different phenomena across different times; the latter damps matter perturbations at and after recombination while the former damps the magnetic fields themselves at much earlier times.} We then plot in Fig.~\ref{fig:PK} the magnetically-induced matter power spectrum across a host of different PMF model parameters. Generally, increasing the magnetic amplitude $\sigma_{B,0}$ increases the PMF-induced matter power spectrum while also shifting its peak towards larger scales. This is as a result of Eq.~(\ref{eq:jeansLength}), where increasing the amplitude of the magnetic field background also increases the magnetic Jeans length; therefore, the density power spectrum need not strictly increase via a vertical shift with increasing $\sigma_{B,0}$. Increasing the magnetic spectral index $n_B$ will increase both the slope and amplitude of the density power spectrum while maintaining its peak at roughly the same scale.

\section{21-cm Signal Formalism} \label{sec:21cm}

A standard probe of cosmic dawn and reionization is the photon emitted from the hyperfine spin-flip transition of an electron in a neutral hydrogen atom. Several effects described below can cause a hydrogen ground-state electron to alternate between its triplet and singlet states, producing a low-energy photon of a wavelength of approximately $21.11 \, \mathrm{cm}$. The early universe was pervaded with neutral hydrogen gas, whose observable features change as the first emitting sources appear; therefore, characterizing and mapping the evolution of the 21-cm background can be an auxiliary probe into high redshift cosmology and astrophysics.

The 21-cm signal is usually quantified via a differential brightness temperature
\begin{equation}
    T_{21} = \frac{T_S - T_\mathrm{rad}}{1+z} \left(1 - e^{-\tau_{21}} \right).
\end{equation}
where $T_\mathrm{rad}$ is the temperature of any background thermal radiation (usually assumed to be strictly the CMB temperature $T_\mathrm{rad} \equiv T_\mathrm{CMB} = 2.7255 \left( 1+z \right)$) \citep{fixsen09}. Here, $\tau_{21}$ is the 21-cm optical depth \citep{furlanetto06b, pritchard08} expressed as
\begin{equation}
    \tau_{21}=\frac{3 h A_{10} c \lambda_{21}^2 n_{\mathrm{H} \mathrm{I}}}{32 \pi k_{\mathrm{B}} T_{\mathrm{S}}(1+z)\left(\partial v_r/ \partial r\right)},
\end{equation}
which is defined in terms of the Einstein A-coefficient $A_{10}$ for 21-cm emission, the radial comoving velocity gradient $\partial v_r/ \partial r$, the wavelength $\lambda_{21}$ of 21-cm radiation, the neutral hydrogen number density $n_\mathrm{HI}$, and fundamental constants. Finally, $T_s$ is the non-thermodynamic spin temperature of the neutral hydrogen in the IGM, which measures the relative occupation numbers of the electrons in either the triplet or singlet hydrogen spin states via $n_1 / n_0 = g_1 / g_0 \, \mathrm{exp} \left( hc / \lambda_{21} k_B T_s\right)$. It can be calculated via the relation
\begin{equation}
    T_s^{-1}=\frac{x_{\mathrm{rad}} T_{\mathrm{rad}}^{-1}+x_c T_k^{-1}+x_\alpha T_c^{-1}}{x_{\mathrm{rad}}+x_c+x_\alpha},
\end{equation}
which is a harmonic mean of $T_\mathrm{rad}$, the color temperature $T_\alpha$ \citep{hirata06}, and the kinetic temperature $T_K$, weighted by the coefficients from Lyman-$\alpha$ (Ly$\alpha$) coupling through the Wouthuysen-Field effect $x_\alpha$ \citep{wouthuysen52, field59}, collisional coupling $x_c$ \citep{loeb04}, and radiation coupling $x_\mathrm{rad} \equiv \left( 1-e^{-\tau_{21}}\right) / \tau_{21}$.

The evolution of the 21-cm \textit{global signal}, defined as the average brightness temperature $\bar{T}_{21}\left(z\right)$ measured against the monopole of the CMB, is subject to the redshift evolution of each aforementioned coupling coefficient \citep{pritchard12}. At ultra-high redshifts $\left( 200 \geq z \geq 30 \right)$, the high gas density of the IGM prompted collisional coupling to dominate; this yields $T_k \propto (1+z)^2$ and a slight absorption signal. As the Universe expands, the IGM density drops, collisional coupling becomes subdominant compared to CMB coupling, and the differential brightness temperature fades. The birth of the first stars at subsequent redshifts $\left( 30 \geq z \geq 20 \right)$ produces an ultraviolet (UV) background; because $x_\alpha$ is proportional to the Ly-$\alpha$ flux $J_\alpha$ \citep{hirata06}, which is itself proportional to the global star formation rate density (SFRD), the onset of Cosmic Dawn (CD) is usually defined as the moment where $x_\alpha$ becomes dominant. Such Ly-$\alpha$ coupling produces an expectedly stronger absorption signal. Continued IGM heating by X-ray \citep{ciardi10, xu14, ewall16, sazonov17} and other energetic photons \citep{chen04, chuzhoy07} from the first sources will eventually result in 21-cm signal in emission marking the beginning of the EoR at $\left( 20 \geq z \geq 6 \right)$. This signal is expected to decay asymptotically as more neutral hydrogen is ionized \citep{benson06, furlanetto06}.

The statistics of 21-cm brightness temperature anisotropies can be quantified to first order through the 21-cm \textit{power spectrum} $P_{21}(\textbf{k},z)$. For convenience, we use the reduced 21-cm power spectrum described by
\begin{equation}
    \Delta_{21}^2(\textbf{k},z)=\frac{k^3 P_{21}(\textbf{k},z)}{2 \pi^2} \, \left[\mathrm{mK}^2\right],
\end{equation}
where 
\begin{equation}
    \left\langle\delta T_{21}(\mathbf{k},z) \delta T^{*}_{21}\left(\mathbf{k}^{\prime},z\right)\right\rangle=(2 \pi)^3 \delta_D\left(\mathbf{k}+\mathbf{k}^{\prime}\right) P_{21}(\mathbf{k},z),
\end{equation}
and $\delta T_{21}(\mathbf{k},z)$ is the Fourier transform of $T_{21}\left(\textbf{x},z\right) - \bar{T}_{21}\left(\textbf{x},z\right)$. As a zero-order approximation, the reduced power spectrum can be thought of as $\Delta_{21}^2(z) \propto \bar{T}^2_{21}(z)$, i.e. the magnitude of the global signal also defines the amplitude of the power spectrum at a specific redshift.

\begin{figure*}[t!]
    \centering
    \includegraphics[width=\textwidth]{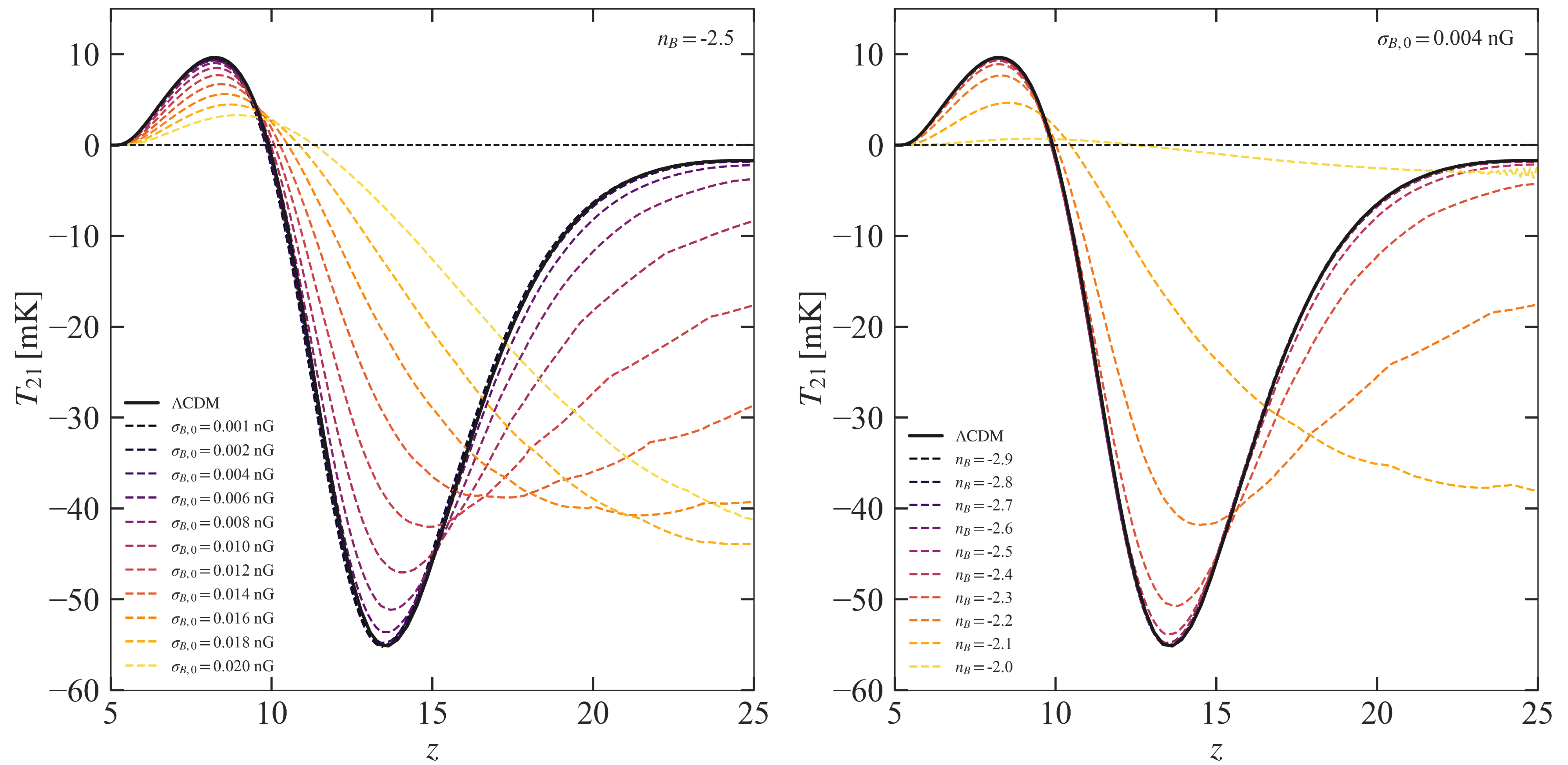}
    \caption{A plot of the 21-cm brightness temperature across redshift, across a variety of different PMF models. Left: Fixed $n_B = -2.5$ and varying magnetic amplitude $\sigma_{B,0}$. Right: Fixed $\sigma_{B,0} = 0.004 \, \mathrm{nG}$ and varying magnetic spectral index $n_B$. An increase in magnetic amplitude results in shallower emission and absorption features that shift toward earlier redshifts. Increasing the spectral index does the same, to a more drastic extent.}
  \label{fig:GS}
\end{figure*}

\section{Simulation} \label{sec:sim}

In this section, we calculate the effects of primordial magnetic fields on density perturbations and their resultant amendments to 21-cm signatures with our aforementioned prescription. We model 21-cm observables by integrating the corresponding $\mathrm{\Lambda CDM}$ + PMF cosmology into an amended version of {\tt\string 21cmFAST v3.1.3} \citep{flitter22, sarkar22, sarkar23} \citep[see original in][]{murray20, mesinger11}. In this modified version, exotic cosmologies from {\tt\string CLASS} and other independently computed matter transfer functions are more flexibly incorporated into the {\tt\string 21cmFAST} excursion set formalism, allowing for beyond-$\mathrm{\Lambda CDM}$ simulations of ionized and neutral hydrogen density fields. \hac{For our analysis, we input our custom matter power spectrum denoted in Eq. \ref{eq:totalMatterPK} including all relevant damping effects. We then employ this change to modify relevant quantities, namely the matter variance $\sigma^2(R, z)$ and the halo mass function $dn / dM$.} Feedback effects from $H_2$-dissociating Lyman-Werner radiation \citep{haiman97,ricotti01, bromm04, ahn09, safranek12, fialkov13, visbal14} and baryon-dark matter relative velocities \citep{tseliakhovich10, yacine14, stacy11, fialkov12, bovy13, fialkov14, barkana16, schmidt16} can also be computed directly from external transfer function inputs. Version {\tt \string 3.1.3} of {\tt \string 21cmFAST} also distinguishes the impact of population-II and -III stars into atomic (ACGs) and molecular cooling galaxies (MCGs) respectively---the latter of which reside in mini-haloes---which contribute different astrophysics into each simulated voxel. We incorporate such effects by using the fiducial astrophysical parameters taken from the \textit{Evolution of 21-cm Structure} project (EOS2021), listed in Table 1 of \citet{munoz22}.  We use the extended Press-Schechter halo mass function and incorporate the effect of mini-haloes in our simulations. The simulated lightcones have sizes 400 \si{Mpc} with a resolution of 4 \si{Mpc} per voxel. 21-cm power spectra were computed from such lightcones using the {\tt\string powerbox} module \citep{murray18}. We adopt the best fit \textit{Planck 2018} TT,TE,EE+lowE+lensing cosmological parameters \citep{planck18}, which assumes $\Omega_m = 0.3153$, $\Omega_b = 0.0493$, $h = 0.6736$, and $n_s = 0.9649$. Additionally, we assume $A_s = 2.035 \times 10^9$, which in {\tt\string 21cmFAST} is the default equivalent to $\sigma_8 = 0.811$ \citep{munoz22}.

\subsection{Global Signal}

We demonstrate in Fig.~\ref{fig:GS} the evolution of the 21-cm global signal over redshift across cosmologies incorporating various PMF parameters, alongside the fiducial global signal prediction generated from a pure $\mathrm{\Lambda CDM}$ cosmology. As the magnetic background amplitude $\sigma_{B,0}$ increases, the absorption feature due to Ly-$\alpha$ coupling translates to higher redshifts. The absorption shift towards earlier times is attributed to the abundance of PMF-generated small-scale matter power which triggers earlier amplified SFRDs dominated by pop-III stars. These enhanced SFRDs translate to an earlier UV background from the first emitting sources which allows Ly-$\alpha$ coupling to dominate the 21-cm spin temperature at higher redshifts. The region of zero brightness temperature that separates the Ly-$\alpha$ and collisional coupling regimes also becomes less pronounced with increasing magnetic field strength, suggesting that the two epochs may become indistinguishable with adequate magnetic amplitude. 

Fig.~\ref{fig:GS} also indicates that, with increasing magnetic background, the PMF-enhanced SFRD tends to attenuate the absorption and emission features in the global signal. The magnitude of the absorption trough is governed predominantly by two competing effects. While higher UV backgrounds cause stronger Ly-$\alpha$ coupling to cold gas and deepens the absorption signal, X-rays from early sources will increase the kinetic temperature of IGM gas. For absorption troughs below $z \lesssim 18$, it is apparent that an $n_B = -2.5$ PMF boosts the cosmic SFRD and causes X-ray heating to predominate over Ly-$\alpha$ coupling. Continued heating at lower redshifts expends the cosmic supply of neutral hydrogen, resulting in less pronounced emission signatures with increasing PMF strength. The global signal across all PMF models fades by redshift $z \sim 5$ as all neutral hydrogen is eventually depleted, in concordance with recent observations of the ionization history \citep{bosman22}.

Similar but more drastic trends appear when increasing the magnetic spectral index. From Fig.~\ref{fig:PK}, increasing $n_B$ both increases the amplitude of matter perturbations and concentrates PMF-induced matter perturbations around a certain wavenumber range. Increasing $n_B$ therefore produces the similar early onset Ly-$\alpha$ coupling and enhanced X-ray heating rates, along with familiarly shallow absorption and emission features appearing in the global signal. However, sufficiently high $n_B$ will concentrate the PMF-induced excess power on small scales, radically increase the SFRD, and extinguishes any 21-cm features by quickly exhausting neutral hydrogen in the IGM.

\subsection{Power Spectrum}

\begin{figure*}[t!]
    \centering
    \includegraphics[width=\textwidth]{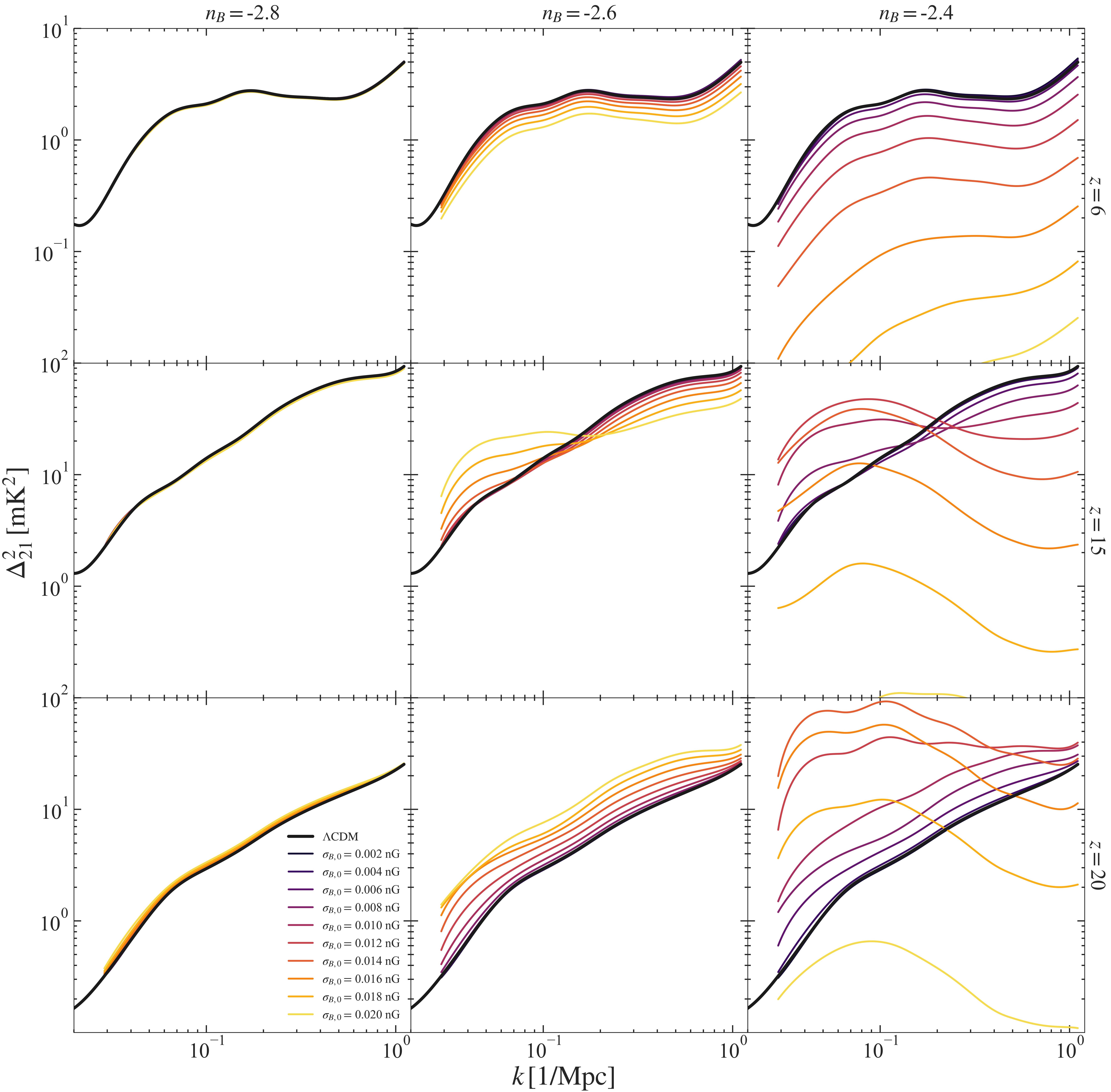}
    \caption{21-cm power spectra across wavenumber at differing redshifts and PMF parameter values. The top, middle, and bottom rows show 21-cm power spectra at redshifts $z=6, 15, 20$ respectively. The left, center, and right columns plot power spectra of PMF models of magnetic spectral indices $n_B = -2.8, -2.6, -2.4$ respectively. Differing magnetic field amplitudes ranging from $0.001 \leq \sigma_{B,0} \leq 0.02 \, \mathrm{nG} $ are depicted in colors of increasing lightness. The 21-cm power spectrum from ordinary $\mathrm{\Lambda CDM}$ without a PMF contribution is overplotted in black for reference. }
  \label{fig:PSvsK}
\end{figure*}

The reduced 21-cm power spectra over wavenumber across different redshifts and magnetic spectral indices are shown in Fig.~\ref{fig:PSvsK}, alongside the fiducial power spectrum generated from a $\mathrm{\Lambda CDM}$ cosmology. As the reduced power spectrum is proportional to the variance of the 21-cm brightness temperature field, the global trend of increasing power at increasing $k$ is expected, especially at higher redshifts. However, when compared to the fiducial model, PMF-inclusive cosmologies induce 21-cm brightness temperature anisotropies whose statistics vary across redshift. 

At high redshifts $(z \sim 20)$, PMF-induced small-scale structure produce large-scale (or low-$k$) temperature inhomogeneities of greater magnitude than the fiducial case. This is attributed to the PMF-enhanced star formation which begins earlier than that from a fiducial cosmology, producing early ionized bubbles that generate large-scale anisotropies. At intermediate redshifts $(z \sim 15)$, the growth of these ionized bubbles leave remnant large-scale anisotropies that exhibit comparatively larger power. However, small-scale anisotropies exhibit lower power compared to the fiducial case; as ionized bubbles expand, neutral hydrogen within bubble cores is exhausted, leading to a suppression at small scales. At lower redshifts close to the end of reionization $z \sim 6$, PMF cosmologies ionize the Universe faster than a fiducial cosmology, leading to roughly global suppression of the temperature power spectrum compared to the $\mathrm{\Lambda CDM}$ prediction. 

These effects are best illustrated with the power spectra generated from a $n_B = -2.6$ cosmology. Towards lower magnetic spectral indices $n_B \lesssim -2.8$, the 21-cm power spectra are nearly identical to the fiducial, due to the corresponding PMF-induced matter power spectrum being subdominant compared to the inflationary spectrum. Models with larger spectral indices $n_B \gtrsim -2.5$ exhibit power spectra with large global suppression. Because increasing the magnetic spectral index also increases the amplitude of the PMF-induced matter power spectrum, progressively higher $n_B$ will cause progressively earlier star formation, resultant depletion of cosmic neutral hydrogen, and an overall weaker signal.

\section{Constraints, Degeneracies and Forecasts} \label{sec:fordeg}
In this section, we deliberate over the plausibility of PMF models in three steps. Firstly, we discern PMF models permitted under external observations. To test the conclusivity of these preliminary restrictions, we assess the assumed astrophysics in our simulations via an information matrix analysis. We then ascertain whether PMF models can be observed by HERA down to these restrictions.

\subsection{External Constraints on 21-cm Observables} \label{sec:ionHistory}

\begin{figure}[t!]
    \centering
    \includegraphics[width=\textwidth/2]{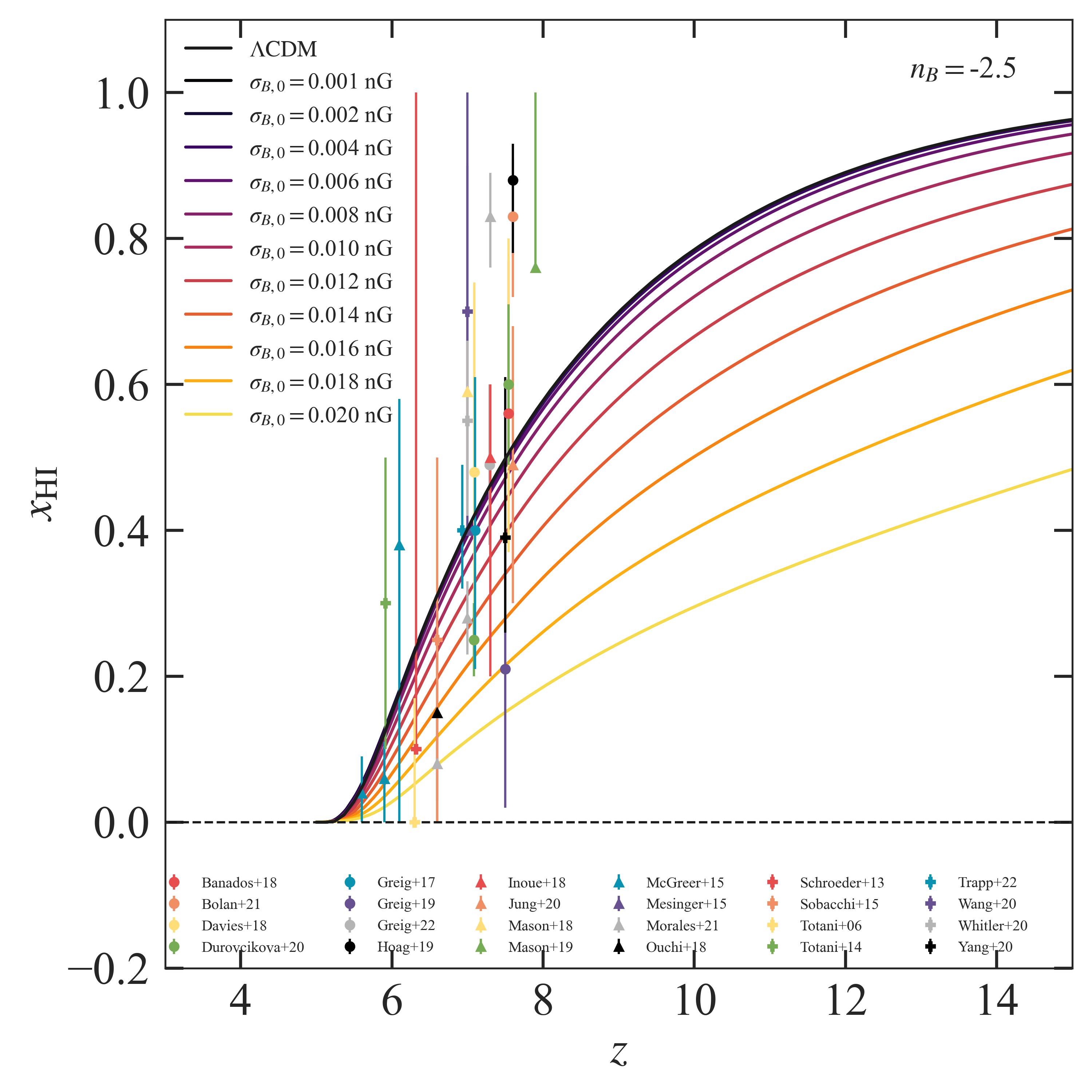}
    \caption{Constraints on PMF models from ionization history, assuming (and without marginalizing over) Planck 2018 cosmological parameters \citep{planck18} and EOS2021 astrophysical parameters \citep{munoz22}. We plot the neutral hydrogen fraction $x_\mathrm{HI}$ evolution over redshift for a $\mathrm{\Lambda CDM}$ + PMF cosmology cosmology of $n_B = -2.5$ and various magnetic amplitudes. Baseline $\mathrm{\Lambda CDM}$ plotted in black. Overplotted are external constraints on the neutral fraction from quasar continuum spectra in the vicinity of the Ly-$\alpha$ line \citep{mason18, durovcikova20, yang20}, quasar damping wings \citep{schroeder13, greig17, banados18, davies18, greig19, wang20, greig22}, from Lyman-break galaxies \citep{mesinger15, hoag19, mason19, jung20, whitler20, bolan22}, Ly-$\alpha$ luminosity functions \citep{inoue18, morales21}, Ly-$\alpha$ emitters \citep{sobacchi15, ouchi18, trapp22}, gamma-ray burst damping wings \citep{totani06, totani14}, and Ly-$\alpha$ and Ly-$\beta$ dark fractions \citep{mcgreer15}.} 
  \label{fig:ionHis}
\end{figure}

Armed with a perturbative formalism to compute the effects of primordial magnetic fields on 21-cm observables, we can identify viable PMF models via their impact on possible astrophysical observables. One promising discriminant of each $\mathrm{\Lambda CDM}$ + PMF cosmology is its ionization history, which is used to compute the optical depth to reionization $\tau_\mathrm{re}$ defined as
\begin{equation}
    \tau_\mathrm{re} = \int \sigma_T n_e(z) dl,
\end{equation}
where $\sigma_T$ is the Thomson scattering cross section of an electron and $dl = c \, dz / H(z) (1+z)$ is the proper cosmological line element. The number density of free electrons $n_e(z)$ can be found by
\begin{align}
    n_e(z) &= \left( n_\mathrm{H}(z) + n_\mathrm{He}(z)\right) \left(1 - x_\mathrm{HI}(z) \right)  \\
    &= \frac{\rho_\mathrm{crit, 0} \Omega_{b,0}(1+z)^3}{4 m_p} \left(Y_\mathrm{He} +4 (1-Y_\mathrm{He}) \right) \left(1 - x_\mathrm{HI}(z) \right) \nonumber
\end{align}
where $\rho_\mathrm{crit, 0} \equiv 3 H_0^2 / 8 \pi G$ is the present-day critical density, $\Omega_{b,0}$ is the present-day baryon density, $m_p$ is the proton mass, and $Y_\mathrm{He}$ is the helium mass fraction. We assume that HI and HeI are ionized at the same rate, and therefore their relative proportions are described by the same neutral fraction $x_\mathrm{HeI}(z) = x_\mathrm{HI}(z) \equiv n_\mathrm{HI}(z) / n_\mathrm{H}(z)$. Due to the relatively high redshifts considered in this work $5 \lesssim z \lesssim 25$, we exclude the negligible contribution from the reionization of the second helium electron (HeII). We compare the PMF-modified neutral fractions across different PMF cosmologies against ancillary observations \citep{mason18, durovcikova20, yang20, schroeder13, greig17, banados18, davies18, greig19, wang20, greig22, mesinger15, hoag19, mason19, jung20, whitler20, bolan22, inoue18, morales21, sobacchi15, ouchi18, trapp22, totani06, totani14, mcgreer15}. These ionization histories are used to compute $\tau_\mathrm{re}$ of each model, from which we seek models in agreement with external constraints on the reionization optical depth. 

Fig.~\ref{fig:ionHis} demonstrates the sensitivity of reionization to sub-nG level fields. PMF models of strength greater than $\sigma_{B,0} \gtrsim 0.014 \, \mathrm{nG}$ for the given $n_B = -2.5$ generate ionization histories that intercept less than half of the available external $x_\mathrm{HI}$ constraints. To more rigorously characterize allowed PMF models, we restrict the PMF parameter space by imposing a $2\sigma $ concordance with the reionization optical depth derived from \textit{Planck 2018} temperature and polarization data, $\tau_\mathrm{CMB} = 0.0627^{+0.0050}_{-0.0058}$  \citep{debelsunce21}. We fit a third-order polylogarithmic function to the error contours and find the PMF parameter space permitted within $2\sigma $ of $\tau_\mathrm{CMB}$ to be
\begin{align} \label{eq:polylogTau}
    \log_{10} \left( \sigma_{B,0}\right) & \leq 7.5 \log_{10}^3 \left|n_B\right| + 3.2 \log_{10}^2\left|n_B\right| \\ \nonumber
    & + 3.3\log_{10}\left|n_B\right| - 4.3 \\ \nonumber
    & \left(\text{for } -2.9 \leq n_B \leq -2.0 \right).
\end{align}

\begin{figure*}[t!]
    \centering
    \includegraphics[width=\textwidth]{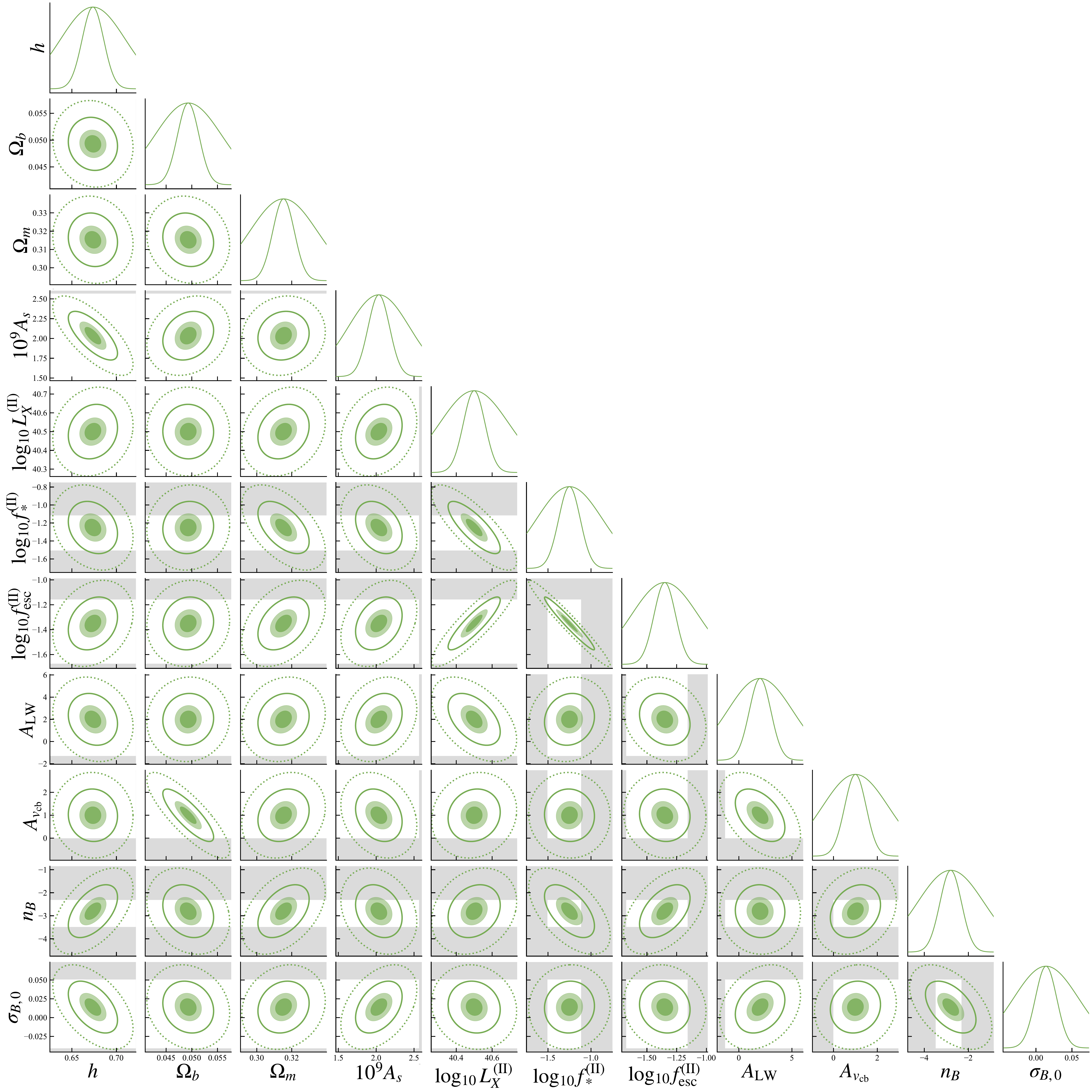}
    \caption{Degeneracies of the PMF parameters for the case $n_B = -2.8, \sigma_{B,0} = 0.014 \, \mathrm{nG}$, alongside the 11 cosmological and astrophysical parameters with the most impact on 21-cm observables as listed in Eq.~(\ref{eq:fisherParams}). We depict the errors from the \textit{long} 6570-hour observation campaign using the shaded contours in opaque (translucent) green to represent is the $1\sigma-$ ($2\sigma-$) errors of each parameter; outlined in solid (dotted) green contours are the $1\sigma-$ ($2\sigma-$) errors for the \textit{short} 730 hour observation campaign. For each plot varying two parameters, we marginalize over the other nine parameters and compute the optical depth to reionization $\tau_\mathrm{re}$ within the parameter space. Shaded in gray are parameter spaces areas excluded by the CMB optical depth ($\tau_\mathrm{re}> \tau_\mathrm{CMB} \pm 2 \sigma_{\tau_\mathrm{CMB}}$)  \citep{debelsunce21}. }
  \label{fig:cornerPlot}
\end{figure*}

We stress that this result should be interpreted as illustrative rather than conclusive. While the Planck 2018 cosmological parameters \citep{planck18} have sub-percent level errors, the EOS2021 astrophysical parameters that we assume in this study \citep{munoz22} are subject to uncertainties stemming from the dearth of observations during cosmic dawn and reionization. To compute these bounds based on optical depth, each simulation fixes and does not marginalize over these parameters. To validate our restrictions on the PMF parameter space, we conduct a more exhaustive analysis of our assumptions subject to $\tau_\mathrm{re}$-based limitations in the following section.

\subsection{Information Matrix Analysis} \label{sec:fisher}

The uncertain astrophysics employed in our simulations imply that our optical depth criteria may result in potentially indeterminate restrictions on PMF models. To mitigate this ambiguity, we challenge the astrophysical parameters themselves. In this section, we examine possible covariances and degeneracies across fiducial \textit{Planck 2018} cosmological and EOS2021 astrophysical parameters.

Observant across a frequency band of $50 \, \mathrm{MHz} \lesssim \nu_\mathrm{obs} \lesssim 250 \, \mathrm{MHz}$, HERA will measure 21-cm fluctuations from cosmic dawn to the end of reionization ($5 \lesssim z \lesssim 27$) across a survey area of $1440 \, \mathrm{deg^2}$. However, a forecast of experimental sensitivities to PMF-inclusive cosmologies is predicated on the uncertainty in measurements of 21-cm power spectra (denoted by $\delta \Delta^2_{21} (k,z)$). To find $\delta \Delta^2_{21} (k,z)$ from a specific interferometric experiment, one must compute the $u-v$ sensitivities of each antenna, characterize possible cosmic variance observed over each baseline, and determine the signal-to-noise of a given observation run time.

We simulate HERA at its completed design performance using {\tt\string 21cmSense} \citep{pober13, pober14} across a bandwidth of $8 \, \mathrm{MHz}$. The frequency range of HERA is divided into 15 frequency bins for our investigation. We assume that HERA is composed of a 350-element interferometer consisting of 14 m parabolic dishes, 320 of which are arranged in a dense hexagonal tessellation and 30 at farther $\sim 0.8 \, \mathrm{km}$ baselines. We simulate two cases of observation runs (hereafter denoted as the \textit{short} and \textit{long} campaigns) in which HERA will conduct either a preliminary observation of two hours per evening over one year, or a full lifetime run of six hours per evening over a duration of three years. We assume a receiver temperature of $T_\mathrm{rec} = 100 \, \mathrm{K}$, a sky temperature of $T_\mathrm{sky} = 60 \, \mathrm{K} (\nu/300 \, \mathrm{MHz})^{-2.55}$. Additionally, we assume a "moderate" foreground scenario, in which foreground contamination in k-space is assumed to extend to $\Delta k_\parallel = 0.1 h \, \mathrm{Mpc}^{-1}$ and baselines are added coherently.

To determine the degeneracies between $\sigma_B, n_B$, and other possible cosmological and astrophysical parameters, we perform an information matrix analysis without priors according to \citep{jungman96, bassett09, jimenez14, munoz20}
\begin{equation}
F_{\alpha, \beta}=\sum_{k, z} \frac{\partial \Delta_{21}^2(k, z)}{\partial \alpha} \frac{\partial \Delta_{21}^2(k, z)}{\partial \beta} \frac{1}{\left[\delta \Delta_{21}^2(k, z)\right]^2},
\end{equation}
where $\alpha$ and $\beta$ represent the varied parameters of our model deemed to have the highest impact on 21-cm observables \citep{sarkar22, flitter22, sarkar23}
\begin{multline} \label{eq:fisherParams}
    (\alpha, \beta) \in   \{ h, \Omega_m, \Omega_b, A_s, \sigma_B, n_B\\
     \log _{10} L_X^{(\mathrm{II})}, \log _{10} f_*^{(\mathrm{II})}, \log _{10} f_{\mathrm{esc}}^{(\mathrm{II})}, A_{\mathrm{LW}}, A_{v_{\mathrm{cb}}} \}.
\end{multline}
Here, the varied cosmological parameters involve the Hubble constant $h \equiv \ H_0 / 100 \ \si{km/s/Mpc}$, the total matter and baryon densities $\Omega_m$ and $\Omega_b$, and the power spectral amplitude $A_s$. Among the varied astrophysical parameters are the SFR-normalized pop-II star X-ray luminosity $\log _{10} L_X^{(\mathrm{II})}$, the pop-II star formation efficiency (SFE) $\log _{10} f_*^{(\mathrm{II})}$, the escape fraction of ionizing pop-II star photons $\log _{10} f_{\mathrm{esc}}^{(\mathrm{II})}$, and the amplitudes of Lyman-Werner (LW) and $v_\mathrm{cb}$ feedback $A_\mathrm{LW}$ and $A_{v_\mathrm{cb}}$ \citep{munoz22}. For consistency, if a pop-II parameter is varied $(\log _{10} L_X^{(\mathrm{II})}, \log _{10} f_*^{(\mathrm{II})}, \log _{10} f_{\mathrm{esc}}^{(\mathrm{II})})$, its analogous pop-III parameter $(\log _{10} L_X^{(\mathrm{III})}, \log _{10} f_*^{(\mathrm{III})}, \log _{10} f_{\mathrm{esc}}^{(\mathrm{III})})$ is also varied.

\begin{figure*}[t!]
    \centering
    \includegraphics[width=\textwidth]{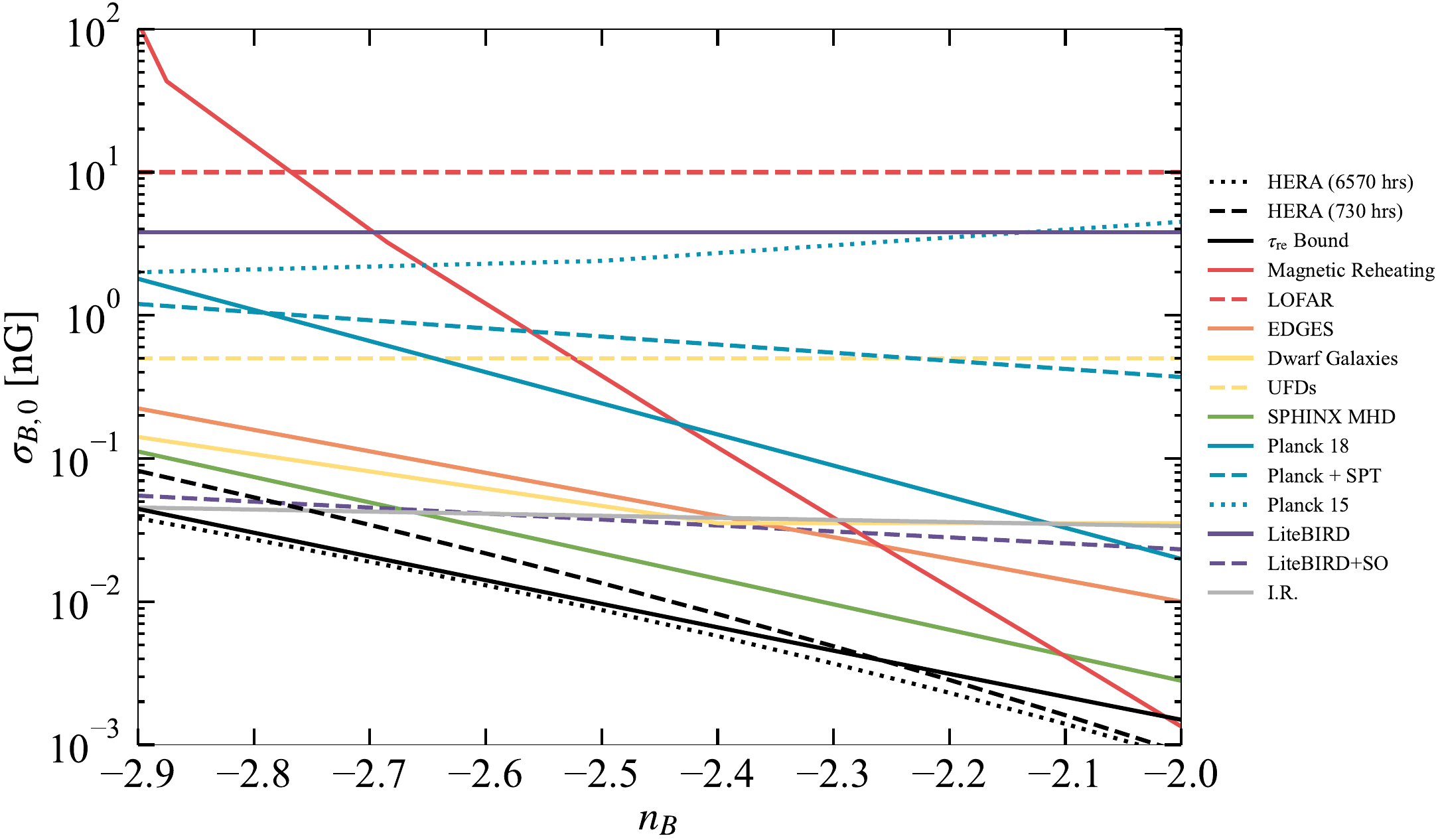}
    \caption{Sensitivity plot of HERA to PMF-inclusive cosmologies with upper bounds from other experiments. We simulate two observation runs, in which HERA records 365 days of observations at two hours per day (dashed black) and 1095 days of observations at six hours per day (dotted black). Here, we use {\tt\string 21cmSense} \citep{pober13, pober14} to determine the experimental signal-to-noise of HERA and forecast measurement errors on the 21-cm power spectrum. Then, for each magnetic spectral index $n_B$, we find where the magnetic amplitude is found to be greater than or equal to twice its experimental uncertainty $(\sigma_{B,0} \geq 2\sigma_{\sigma_{B,0})}$. We marginalize over all cosmological and astrophysical parameters to yield errors strictly for the PMF parameters. $\tau_\mathrm{CMB}$ constraints from this work (dotted black) plotted for reference. Overplotted are the upper bounds from different probes and experiments: from magnetic reheating constraints derived from big-bang nucleosynthesis + the CMB \citep{saga18} (solid red), from IGM accretion shocks observed through LOFAR \citep{locatelli21} (dotted red), from the 2018 EDGES detection \citep{minoda19} (orange), from simulations of dwarf galaxies \citep{sanati20} (solid yellow), from observations of ultra-faint dwarf galaxies \citep{safarzadeh19} (dashed yellow), from $\tau_\mathrm{re}$ constraints from the SPHINX-MHD simulations \citep{katz21} (green), from Planck 2018 polarization data \citep{paoletti22} (solid blue), from combined Planck-South Pole Telescope measurements \citep{zucca17} (dashed blue), from earlier Planck 2015 data \citep{planck15} (dotted blue), from the upcoming LiteBIRD and Simons Observatory experiments \citep{paoletti19} (solid and dashed violet), and from CMB studies of inhomogeneous recombination \citep{jedamzik19, galli22} (gray). \hac{Our results demonstrate two important points. Firstly, our HERA sensitivities are lower than the sensitivities of current 21cm observatories and future standalone CMB experiments. Secondly, HERA at its longest observation run will be sensitive to models even below the imposed (unmarginalized) optical depth bound.} }
  \label{fig:21sensitivity}
\end{figure*}

Shown in Fig. \ref{fig:cornerPlot} is the result of our forecast for both \textit{short-} and \textit{long-}observation runs. Here, we use a PMF model of $n_B = -2.8, \sigma_{B,0} = 0.014 \, \mathrm{nG}$ which predicts an optical depth of $\tau_\mathrm{re} = 0.0642$. Since the predicted $\tau_\mathrm{re}$ for this PMF model is safely within the CMB bounds, this specific information matrix analysis can be taken without loss of generality and is applicable to other concordant PMF-inclusive cosmologies. Areas shaded in gray are disfavored by the CMB optical depth ($ |\tau_\mathrm{re} - \tau_\mathrm{CMB} | \geq 2\sigma_{\tau_\mathrm{CMB}}$.) Comparatively, the cosmological parameters do not have as strong a $\tau_\mathrm{re}$ constraint though each parameter is known to sub-percent level precision; only the amplitude of perturbations $A_s$ shows a slight upper bound. From the astrophysical parameters, the star formation efficiency $\log _{10} f_*^{(\mathrm{II})}$ enjoys the strongest bounds. This makes sense, as a higher (lower) SFE results in more (less) luminous sources that dramatically hastens (delays) reionization. \hac{Though the \textit{short}-campaign observation of $f_*$ shows a covariance overlapping regions of excluded parameters, Fig. \ref{fig:cornerPlot} demonstrates that a \textit{long}-campaign run is sufficient to resolve parameters from excluded cosmologies.} \hac{The \textit{long-}campaign covariance of the escape fraction $\log _{10} f_{\mathrm{esc}}^{(\mathrm{II})}$ shows it is allowed to vary at most from $\sim 3.5-5.5\%$ for Pop II haloes of mass $10^{10} \, M_\odot$ and Pop III haloes of mass $10^7 \, M_\odot$. This constraint on $\log _{10} f_{\mathrm{esc}}^{(\mathrm{II})}$ is reassuringly stronger than its $\tau_\mathrm{re}$ upper bound. Though an increase in escaping ionizing photons results in faster reionization, our analysis suggests that a \textit{long-}campaign will show that $\log _{10} f_{\mathrm{esc}}^{(\mathrm{II})}$ can vary less than what concordance cosmology can permit.} Of the PMF parameters, $n_B$ shows strong constraints while $\sigma_{B,0}$ has only a strong upper bound. This is because the magnetic spectral index both influences the PMF-induced matter power spectrum amplitude and the concentration of matter power across scale, which from Fig.~\ref{fig:GS} is known to shift the time of reionization more effectively than the magnetic amplitude.

Our analysis suggests that the \textit{long-}campaign parameter covariances from a full HERA lifetime run are confidently within optical depth bounds established independently from this work. However, our degeneracy inquiry may not be fully conclusive. For every plot in Fig.~\ref{fig:cornerPlot}, we vary two parameters and impose $\tau_\mathrm{re}$ constraints only after having marginalized over the other nine. Nonetheless, our investigation represents an important first step in distinguishing the correct astrophysics represented by confidently chosen parameters. A fully rigorous analysis involves mapping the eleven-parameter space without marginalization, which is computationally intensive for a simulation-based information matrix forecast and will likely require an emulator-based approach~\cite{lazare23}. We leave this analysis for future work.

\subsection{21-cm Experimental Sensitivities} \label{sec:hera}

In previous sections, we found an optical depth restriction to the PMF parameter space with respect to a set of unmarginalized fiducial astrophysical parameters, and then performed an information matrix analysis of our parameters subject to these restrictions. \hac{The latter analysis demonstrated that our choice of astrophysical parameters have little bias our optical depth computations; only magnetic field strength and spectral index can appreciably modify $\tau_\mathrm{re}$.} We now inquire whether HERA can observe PMFs within the framework of these constraints. To derive the sensitivity of HERA to PMFs, we now add Planck 2018 \citep{planck18} priors to the information matrix and take the inverse. Across a host of PMF parameters, we marginalize over all other non-PMF parameters and find $\sigma_{\sigma_{B,0}}$, the 1-$\sigma$ error on the smoothed magnetic amplitude $\sigma_{B,0}$. We then consider the PMF parameter space which can be detected. For a given $n_B$, we characterize detectability as the condition under which the magnetic amplitude can be observed to twice its error or greater $(\sigma_{B,0} \geq 2\sigma_{\sigma_{B,0}})$. We portray the forecasted HERA sensitivity in Fig.~\ref{fig:21sensitivity}, described by the best-fit polylogarithmic function 
\begin{align}
    \log_{10} (\sigma_{B,0}) &= a \log_{10}^2 |n_B| + b \log_{10} |n_B| - c \nonumber \\ 
    & \left(\text{for } -2.9 \leq n_B \leq -2.0 \right). 
\end{align}
in which $(a, b, c) = (-0.2, 12.3, -6.7)$ for the \textit{short} campaign, and $(a, b, c) = (-4.0, 13.4, -6.6)$ for the \textit{long} campaign. We also portray the functional form of the unmarginalized upper bound from $\tau_\mathrm{CMB}$ in Eq.~\ref{eq:polylogTau}. For comparison, we also plot existing constraints and forecasted bounds from current and upcoming observations and experiments \citep{saga18, minoda19, sanati20, safarzadeh19, katz21, paoletti22, zucca17, paoletti19, locatelli21}.

As anticipated, HERA sensitivities improve with increasing magnetic spectral index. This is because larger $n_B$ also increases the amplitude of the PMF-induced matter power spectrum in addition to its slope. Therefore, for a given magnetic amplitude, more excess small scale structure is expected for higher magnetic spectral indices than lower ones. Given that the 21-cm power spectrum is roughly proportional to the square of the global signal, and that excess small-scale power generally shifts the 21-cm global signal towards earlier times, it is remarkable that the sensitivity improves with increasing $n_B$ even with a fainter, more distant signal. 

\hac{Our results demonstrate two important points of note. Firstly, our upper bounds and sensitivities for both \textit{short-} and \textit{long-} campaigns are lower than the bounds from the most competitive of MHD \citep{katz21} and dwarf galaxy \citep{sanati20} simulations, and are more sensitive toward PMF cosmologies than the next generation of standalone experiments \citep{planck15, zucca17, minoda19, paoletti19, paoletti22}. Towards scale-invariant PMF power spectra $n_B \approx -3$, our forecasted sensitivities for the three-year observation run are compatible but still below that from the combined LiteBIRD + SO polarization data \citep{paoletti19} and constraints from studies of inhomogeneous recombination \citep{jedamzik19, galli22}, which get progressively stronger with increasing magnetic spectral index. Secondly, our forecasts show that a three-year observation run with HERA can probe PMF cosmologies with parameters below our optical depth bound. Though different assumptions of fiducial astrophysical parameters could alter the predicted sensitivity, our analysis in Sec. \ref{sec:fisher} shows that these parameter-dependent modifications to 21cm observables should be subdominant to those modified by the magnetic field strength and spectral index. Since our choice of parameters are still safely within the strict confines of the $\tau_\mathrm{re}$ limits in Fig. \ref{fig:21sensitivity}, our assumption-dependent sensitivity calculation is still a credible first estimate.} HERA should therefore be powerful in constraining $\sim \mathrm{pG}$ level cosmic fields, far below that of the next generation of cosmological experiments.

\section{Discussion \& Conclusion} \label{sec:conclusion}

Due to the paucity of observations during cosmic dawn and reionization, many questions on the origin and evolution of cosmic magnetic fields persist. In this work, we have shown how primordial magnetic fields can be probed and constrained with 21-cm cosmology. In our prescription \citep{magFieldPaper}, we show how PMFs subject to Alfv\'en damping and Jeans suppression effects can generate matter density perturbations at small scales. We integrate this prescription into a modified version of {\tt\string 21cmFAST v3.1.3} to compute 21-cm observables. Our results reaffirm that more small-scale matter power will augment global star formation rates, driven by pop-III stars at early times and pop-II stars at later times. The resultant excess radiation background will accelerate reionization, marked by the position of the 21-cm absorption trough shifted towards higher redshifts. In the 21-cm global signal, we observe consequential X-ray heating to predominate over Ly-$\alpha$ coupling, which culminates in increasingly shallower emission and absorption signals with increasing PMF strength. The statistics of 21-cm fluctuations also alter accordingly, leading to altered brightness temperature power spectra from the abundance of early ionized bubbles.

As a preliminary constraint, we impose existing bounds on the optical depth to reionization to find the space of allowed PMF models. We compute a grid of simulations across various magnetic amplitudes $\sigma_{B,0}$ and spectral indices $n_B$, and find which models fall within a $2\sigma$ of the optical depth predicted from CMB polarization. We find that, for PMF models of $n_B \lesssim -2.7$, magnetic amplitudes of order $\sim 10 \, \mathrm{pG}$ are allowed, which falls to $\sim 1 \, \mathrm{pG}$ towards $n_B \gtrsim -2.5$. \hac{Such low constraints on PMF strengths are somewhat below the $10-50 \ \mathrm{pG}$ needed to relieve the Hubble Tension \citep{koh21}.} As the $\tau_\mathrm{CMB}$ criterion is imposed over simulations that are both dependent on and do not marginalize over uncertain astrophysical parameters, this bound is more illustrative than conclusive.

For a more determinate inquiry, we conduct an information matrix analysis on the cosmological, astrophysical, and PMF parameters used in our simulations. Assuming a 730 hour "\textit{short}" and 6570 hour "\textit{long}" observation campaign and imposing no priors, we analyze the degeneracies of each parameter subject to the constraints of the CMB optical depth. We find that $\tau_\mathrm{CMB}$ most restricts the allowed astrophysical parameter space relative to that of the cosmological parameters. We conclude that a \textit{long} observation run over the course of the lifetime of HERA can constrain uncertain astrophysics within the allowed parameter space. A rigorous eleven-parameter analysis of confidence hyperellipsoids compatible with $\tau_\mathrm{CMB}$ is beyond the scope of this work. Nevertheless, our preliminary constraints are a meaningful first step in mapping out the degenerate astrophysics in the early Universe.

Lastly, we forecast the sensitivity of HERA to PMF-inclusive cosmologies by imposing Planck 2018 \citep{planck18} priors and marginalizing over all other parameters. \hac{We find that, over an observation run of three years, HERA is sensitive towards PMFs by nearly an order of magnitude below that of existing 21cm observatories and future standalone CMB experiments.} Though our preliminary $\tau_\mathrm{re}$ bound is dependent on unmarginalized astrophysical parameters, we find this constraint to be compatible with HERA sensitivity. 

Our current study is limited to the most dominant structure-forming influences of PMFs. Secondary effects of possible significance include several magnetohydrodynamical processes \citep{gnedin00a, gnedin00b} such as ambipolar diffusion and MHD turbulence decay, both of which could provide extra heating to the IGM \citep{sethi05, sethi08} and modify the cosmic average ionization rate \citep{schleicher08, schleicher11, chluba15}. Should these heating mechanisms become important, the excess thermal energy can hinder the collapse of early structure, which in turn delay the formation of the first ionizing sources. PMFs can also interfere with the cooling of molecular hydrogen within early haloes, potentially affecting the initial mass function of the first stars \citep{koh21}. More conspicuously, IGM heating will weaken the 21-cm signal during the absorption epoch \citep{tashiro06a}; higher gas temperatures will leave less room for the spin temperature to depart from the background CMB temperature, which will hamper detectability and may weaken our constraints. These counteractive effects imply that our work may overestimate the actual modifications of 21-cm observables by PMFs. \hac{Though it is difficult to ascertain without a full analysis how PMF constraints may shift when both enhanced structure and MHD heating are employed, analyses employing strictly heating mechanisms have shown weaker PMF constraints. Heating-exclusive effects on \textit{y}-type CMB distortions \citep{kunze15} posit $95\%$ bounds to the magnetic amplitude to around $\sigma_\mathrm{B,0} \sim 0.11 - 0.17 \ \si{nG}$, over an order of magnitude less constraining than our analysis. Nevertheless, such secondary repercussions may be pivotal in distinguishing PMFs from other beyond-standard model physics whose consequences on small-scale structure could be degenerate with our work.}

By navigating through cosmic dawn and the epoch of reionization, 21-cm cosmology will introduce powerful insights on exotic deviations from concordance cosmology. HERA is poised to find evidence of primordial magnetic fields with unprecedented sensitivity over current and next-generation experiments. Such discoveries could shed new light on the formation of structure and the complex stellar and extragalactic astrophysics during these uncharted epochs of cosmology.

\acknowledgments

We thank Nashwan Sabti, James Davies, Selim Hotinli, Julian Mu{\~n}oz, Roberto Cotesta, and Andrea Antonelli for useful discussions, and Debanjan Sarkar for his contribution to the modified {\tt\string 21cmFAST v3.1.3} code. HAC was supported by the National Science Foundation Graduate Research Fellowship under Grant No.\ DGE2139757.  TA was supported by a Negev Ph.D. Fellowship awarded by the BGU Kreitmann School. JF was supported by the Zin fellowship awarded by the BGU Kreitmann School. EDK~acknowledges support from an Azrieli faculty fellowship. This work was supported at Johns Hopkins by NSF Grant No.\ 2112699 and the Simons Foundation. This work was carried out at the Advanced Research Computing at Hopkins (ARCH) core facility  (arch.jhu.edu), which is supported by the National Science Foundation (NSF) grant number OAC1920103.

\bibliography{refs}

\end{document}